\documentclass{article}

\usepackage{arxiv}

\usepackage[utf8]{inputenc} 
\usepackage[T1]{fontenc}    
\usepackage{hyperref}       
\usepackage{url}            
\usepackage{booktabs}       
\usepackage{amsfonts}       
\usepackage{nicefrac}       
\usepackage{microtype}      

\usepackage{graphicx}
\usepackage{algorithm}
\usepackage{algpseudocode}
\usepackage{xcolor}
\usepackage{amssymb}
\usepackage{subcaption}
\usepackage{booktabs}
\usepackage{enumerate}
\usepackage{amsmath}
\usepackage{appendix}

\title{Deep Markov Chain Monte Carlo}

\author{
  Babak Shahbaba\thanks{babaks@uci.edu} \\
  Department of Statistics\\
  UC Irvine, CA, USA\\
   \And
 Luis Martinez Lomeli \\
  Mathematical, Computational, and Systems Biology\\
  UC Irvine, CA, USA\\
  \AND
  Tian Chen \\
  Department of Statistics \\
  UC Irvine, CA, USA \\
  \And
  Shiwei Lan \\
  School of Mathematical and Statistical Sciences \\
  Arizona State University, AZ, USA \\
}

\begin{document}
\maketitle

\begin{abstract}
We propose a new computationally efficient sampling scheme for Bayesian inference involving high dimensional probability distributions. Our method maps the original parameter space into a low-dimensional latent space, explores the latent space to generate samples, and maps these samples back to the original space for inference. While our method can be used in conjunction with any dimension reduction technique to obtain the latent space, and any standard sampling algorithm to explore the low-dimensional space, here we specifically use a combination of auto-encoders (for dimensionality reduction)  and Hamiltonian Monte Carlo (HMC, for sampling). To this end, we first run an HMC to generate some initial samples from the original parameter space, and then use these samples to train an auto-encoder. Next, starting with an initial state, we use the encoding part of the autoencoder to map the initial state to a point in the low-dimensional latent space. Using another HMC, this point is then treated as an initial state in the latent space to generate a new state, which is then mapped to the original space using the decoding part of the auto-encoder. The resulting point can be treated as a Metropolis-Hasting (MH) proposal, which is either accepted or rejected. While the induced dynamics in the parameter space is no longer Hamiltonian, it remains time reversible, and the Markov chain could still converge to the canonical distribution using a volume correction term. Dropping the volume correction step results in convergence to an approximate but reasonably accurate distribution. The empirical results based on several high-dimensional problems show that our method could substantially reduce the computational cost of Bayesian inference. 
\end{abstract}

\section{Introduction}
While Bayesian methods can provide a principled and robust framework for data analysis, they tend to be computationally intensive since Bayesian inference usually requires the use of Markov Chain Monte Carlo (MCMC) algorithms to simulate samples from intractable distributions. Although simple sampling methods, such as the Metropolis algorithm, are often effective at exploring low-dimensional distributions, they can be very inefficient for complex and high-dimensional models. In this paper, we propose a computationally efficient algorithm for Bayesian inference in high-dimensional problems. Our approach maps the original parameter space to a low-dimensional latent space, which can be explored efficiently using standard sampling algorithms. The resulting samples are then mapped back to the original parameter space for inference. While theoretically our method can be set up as a proper MCMC algorithm that converges to the true distribution, in practice however, it might be more efficient to trade some accuracy for computational speed by setting up the algorithm such that it converges to an approximate distribution. In this sense, our method shares some similarity with variational Bayes as compared to MCMC.

\paragraph{Recent advances in sampling algorithms} Many computationally efficient sampling algorithms based on geometrically motivated methods, such as Hamiltonian Monte Carlo (HMC) and its variants, have been proposed in recent years. See for example \cite{shahbabaSplitHMC,Welling11,hoffman11,lanICML14,lanAAAI14}. However, to make such geometrically motivated methods practical for big data analysis, one needs to combine them with efficient and scalable computational techniques. One common approach is subsampling \cite{Welling11,hoffmann10,shahbabaSplitHMC, chen14}, which restricts the computation to a subset of the observed data. This is based on the idea that big datasets contain a large amount of redundancy so the overall information can be retrieved from a small subset. In general applications, however, we cannot simply use random subsets for this purpose: the amount of information we lose as a result of random sampling leads to non-ignorable loss of accuracy, which in turn has a substantial negative impact on computational efficiency \cite{betancourt15}. Therefore, in contrast to subsampling, several recent methods have been proposed based on exploring smoothness or regularity in parameter space in order to find detailed and free-form approximations of the target posterior distribution \cite{strathmann2015,Zhang2017a, Zhang2017b, song17, levy2017generalizing, Li2019}.  

\paragraph{Variational Bayes as an alternative to MCMC} When set up properly, an MCMC algorithm can converge to the true target distribution in theory. However, for complex and high dimensional models, waiting for convergence to the exact target distribution might not be practical. A main alternative to MCMC is variational Bayes (VB) inference \cite{jordan99,wainwright08, Honkela10, saul96, salimans13}, which transforms Bayesian inference into an optimization problem where a parametrized distribution is introduced to approximate the target posterior distribution by minimizing the Kullback-Leibler (KL) divergence with respect to the variational parameters. Compared to MCMC methods, VB introduces bias but is usually faster. 

\paragraph{The best of both worlds} It is reasonable to think that a combination of both methods might be able to mitigate their shortcomings. An early attempt in this direction was the work of \cite{Freitas01}, where a variational approximation was used as proposal distribution in a block Metropolis-Hasting (MH) algorithm in order to capture high probability regions quickly, thus facilitating convergence. More recently, some new methods have been proposed that rely on combining fast variational methods with exact MCMC simulations in order to improve the overall accuracy and computational efficiency of Bayesian models applied to big data problems \cite{Salimans15, Zhang2018}

Here, we explore an alternative approach based on finding a low-dimensional representation of the parameter space. The idea is that the seemingly high-dimensional parameter could in fact reside in a low-dimensional subspace. For example, a regression model could include features that are either redundant or unrelated to the response variable. To this end, we propose a novel HMC algorithm, which handles this situation naturally by performing dimensionality reduction in the parameter space.

\paragraph{Outline} Our paper is organized as follows. We first review some existing algorithms and preliminary concepts related to our proposed method. More specifically, we focus on computational challenges of MCMC in high dimensional problems. We then describe our method, \emph{Auto-encoding HMC (AE-HMC)}, in details. We prove that our sampling method can in principle produce a proper Markov chain that converges to the true target distribution. However, we will also show that it is possible to use our method to obtain a reasonably well approximate distribution without a substantial sacrifice in performance. Finally, we examine our method using simulated and real data.

\section{Preliminaries}
A central task of Bayesian inference is to calculate the integral: $$E_{\pi}(f) = \int f(q)\pi(q)dq,$$ where $\pi(q) = p(q)p(D\vert q)$ is the posterior distribution with respect to parameter $q$. The integral is typically high dimensional and intractable. Therefore, we usually resort to numerical methods by obtaining samples from $\pi(q)$ and calculating a finite sum as an approximation to the integral. An accurate approximation usually relies on efficient exploration of typical set, the region in the parameter space which contributes most to the integral \cite{betancourt2017conceptual}. 

\subsection{Pathological Behavior of Random Walk Metropolis in High dimensional space}
By far, the most commonly used sampling method is Markov chain Monte Carlo (MCMC). MCMC method samples from the parameter space by generating a Markov chain which eventually converges to the target distribution (the posterior distribution in Bayesian framework) as its stationary distribution. A new state is proposed at each iteration according to a transition map $T(q^* \vert q)$. In particular, Metropolis-Hastings algorithm is used to construct such a Markov chain by proposing a new state and accepting it with the following probability: $$a(q, q^*) = \min (1, \dfrac{\pi(q^*) T(q \vert q^*)}{\pi(q) T(q^* \vert q)}),$$ which can guarantee the convergence to $\pi(q)$ due to detailed balance. Random walk Metropolis (RWM) is one of the most widely used Metropolis algorithms, with $T(q^* \vert q)$ set to be a Gaussian distribution centered at current state $q$.

Though RWM is simple to implement, it does not scale well to high-dimensional problems --- exploration of typical set in high dimensional space is very challenging for random walk proposals. As discussed by \cite{betancourt2017conceptual}, the region outside the typical set has vanishing densities and large volume, which does not contribute substantially to the integral. Thus, it might not be efficient to spend computational resources to explore this area. As the dimension of the space grows, the volume of the outside region grows exponentially, and overwhelms the volume of the interior region. As a result, RWM tends to propose a state outside the typical set that is rejected with a high probability. 

\subsection{Hamiltonian Monte Carlo}
Faster exploration can be obtained using, for example, Hamiltonian Monte Carlo (HMC), which was first introduced by \cite{duane87} and later reviewed by \cite{neal11}. HMC reduces the random walk behavior of Metropolis by taking $L$ steps of size $\epsilon$ guided by Hamiltonian dynamics, which uses gradient information, to propose states that are distant from the current state, but nevertheless have a high
probability of acceptance. In particular, HMC introduces a set of auxiliary variables $p$ (called momentum) with the same dimension as the original parameters $q$. The parameter space is then augmented to a phase space $(q, p)$, and HMC proposes new states jointly for $(q, p)$, according to Hamilton's equations:
\begin{align*}
	\dfrac{dq_i}{dt} & = \dfrac{\partial H}{\partial p_i} \\
	\dfrac{dp_i}{dt} &= -\dfrac{\partial H}{\partial q_i} 
\end{align*}
where $H = H(q, p) = U(q) + K(p)$. $U(q)$ is associated with the target density, and $K(p)$ is usually chosen to associate with the density of a zero-mean Gaussian with covariance $M$.
\begin{align*}
	U(q) & = - \log \pi(q) = -\log [p(q)p(D\vert q)]\\
	K(p) &= p^TM^{-1}p/2 
\end{align*}

A new state will be proposed from $(q(t), p(t))$ to $(q(t+s), p(t+s))$. Since the Hamiltonian equations are not analytically solvable in general, in practice, we resort to leapfrog method by discretizing the time to approximate the dynamics. Given a step size $\epsilon$ and number of steps $L$, $s$ is defined to be $\epsilon L$. 

Hamilton's equations describe the dynamics of a physical system with conservative energy $H(q, p)$. $q$ is the position of the object, and $p$ is the momentum. Correspondingly, $U(q)$ is the potential energy, and $K(p)$ is the kinetic energy. While $U(q)$ and $K(p)$ are varying as the object moves, the Hamiltonian $H(q,p) = U(q) + K(p)$ is constant.

In its MCMC application, $\exp(-H(q,p))$ corresponds to the joint probability of $(q,p)$, also referred to as canonical distribution:
\begin{align*}
	\pi(q, p)  = \dfrac{1}{Z} \exp(-H(q,p)/T)
\end{align*}  where $Z$ is a normalization constant, and $T$ represents the temperature of the system. Eventually the Markov chain will converge to the canonical distribution due to the reversibility and volume preservation properties of the Hamiltonian dynamics. The marginal distribution of $q$ is exactly the target density.

\subsection{Auto-encoder}
\begin{figure}[!t]
	\centering
	\includegraphics[width=0.4\textwidth]{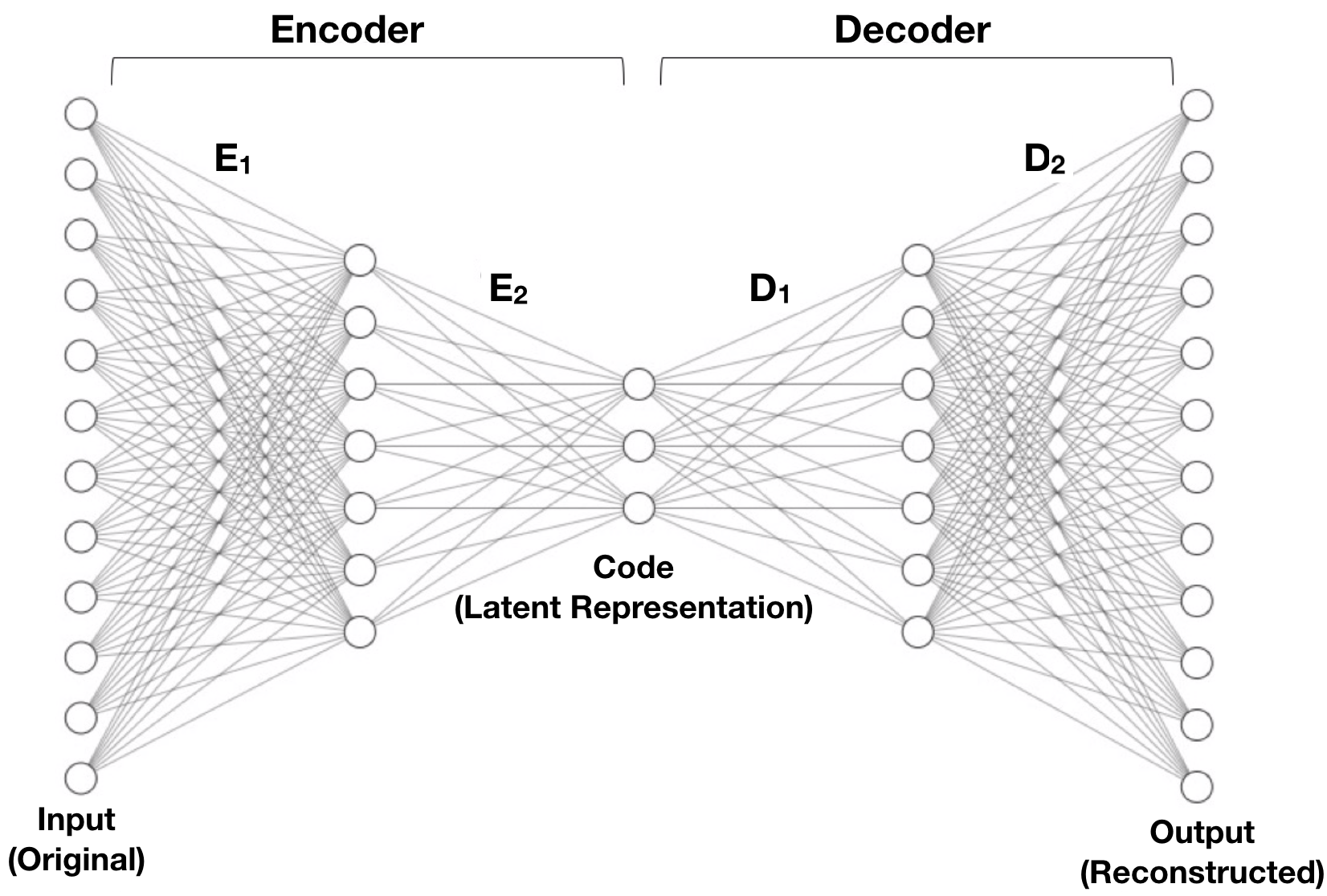}
	\caption{Auto-encoder Network Architecture}
	\label{fig:auto-encoder}
\end{figure}

Auto-encoder is a special type of feed forward neural network for learning latent representation of the data (Figure \ref{fig:auto-encoder}). The data are fed from the input layer and encoded into a low-dimensional latent representation (code). The code is then decoded into a reconstruction of the original data. The goal of auto-encoder is to learn an identity map such that the output (reconstruction) is closely matched with the input data. The model is trained to minimize the difference between the input and the reconstruction. Auto-encoder could learn complicated nonlinear dimensionality reduction and thus is widely used in challenging tasks such as image recognition and artificial data generation. 

According to universal approximation theorem \cite{csaji2001approximation}, a feed-forward artificial neural network can approximate any continuous function given some mild assumptions about the activation functions. Theoretically, an auto-encoder with suitable activation functions could represent an identity map. Therefore, auto-encoder could learn a encoder $\phi$ and decoder $\psi$ such that $\psi \circ \phi = I$. An accurate reconstruction of the data implies a good low-dimensional representation.

\section{Auto-encoding HMC}
While HMC explores the parameter space more effectively, each iteration is computationally demanding since we have to evaluate a high-dimensional gradient function. To alleviate this issue, we propose a new method called Auto-encoding HMC (AE-HMC). First, we collect a small set of posterior samples from the target distribution by running the standard HMC algorithm. We then use the collected samples to find a low-dimensional latent parameter space by training an auto-encoder. Given an initial state in the original parameter space, we use the encoding part of the auto-encoder to find its projection in the latent space, simulate Hamiltonian dynamics to generate a new state, and use the decoding part of the auto-encoder to project it back to the original space to obtain a proposal. While this exploration will not be as accurate as the standard HMC, it could reduce the overall computational cost. 

\paragraph{Illustration} We illustrate our approach using a three-dimensional Gaussian distribution. Because the original dimension is very low, we use a Principle Component Analysis (PCA), which can be considered as a special case of auto-encoder \cite{baldi89}: the encoder of an auto-encoder reduces to a PCA if all the activation functions are linear and the inputs are normalized. Suppose we are interested in sampling from a three-dimensional Gaussian distribution with zero mean and covariance 
$$\Sigma = \begin{bmatrix} 
1 & 0.95 &  0.7 \\ 
0.95 &  1 & 0.5 \\
0.7 &  0.5 & 1 \end{bmatrix}$$
To perform dimension reduction using PCA, we simply find orthornormal matrix $P$ such that $\Sigma' = P \Sigma P^T$ is diagonalized, where the diagonal entries of $\Sigma'$ are the variances of the transformed variables. Here we extract the first two principal components with greatest variance as the low-dimensional representation, and simulate Hamiltonian dynamics in the latent space. 

\begin{figure}[!htb]
		\centering
		\includegraphics[width=0.48\textwidth]{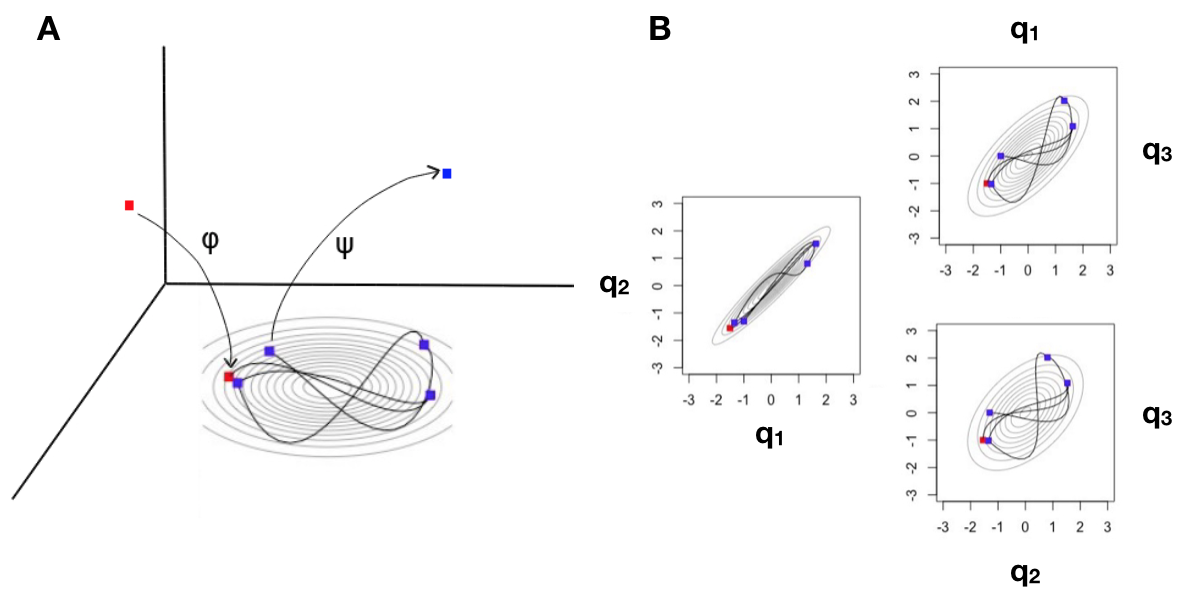}
		\caption{\textbf{A:} HMC trajectory in the latent space (2-dimensional); the red square is the initial position, and the blue squares are HMC proposals. \textbf{B:} Trajectories projected back to original parameter space (3-dimensional) showing that our method can still explore the parameter space effectively.}
		\label{fig:illust}
\end{figure}

As shown in Figure \ref{fig:illust}, we only performed HMC in the space of two dimensions. But when the proposals are projected back to the original space, the algorithm still efficiently explores the space with distant proposals. 

When it comes to high-dimensional problems, we will use auto-encoder for dimension reduction. More specifically, denote the parameters of interest $q_v$ and its latent representation $q_h$. We also denote the encoder and decoder as
\begin{align*}
	\phi &:  q_v \mapsto q_h \\
	\psi &: q_h \mapsto q_v'
\end{align*}
where $q_v'$ is a reconstruction of $q_v$. If the error of the auto-encoder goes to zero, we have:
\begin{align*}
	\psi = \phi^{-1} &: q_h \mapsto q_v
\end{align*}
Our algorithm is composed of the following three stages:
\begin{enumerate}
	\item Pre-sample a few (e.g. $1000$) samples of $q_v$ using standard HMC
	\item Train an auto-encoder to fit the samples, and obtain the fitted encoder $\phi$ and decoder $\psi$
	\item Run AE-HMC to propose $q_v^*$ from $q_v$ (a detailed version is provided in Algorithm \ref{alg:auto-encoder_hmc}):
	\begin{enumerate}[i]
		\item Find $q_h = \phi(q_v)$
		\item Propose $q_h^*$ from $q_h$ by running HMC in the latent space
		\item Obtain $q_v^* = \psi(q_h^*)$
	\end{enumerate}
\end{enumerate}

\subsection{HMC in the latent space}
Denote the complementary momentum of $q_v$ as $p_v$. The corresponding latent parameters are denoted as $(q_h, p_h)$. The auxiliary variable in the latent space is constructed using the same encoder $p_h= \phi(p_v)$, with $p_v$ sampled from a Gaussian distribution. Further, denote the target density $\pi_{q_v}(q_v)$. We set the potential energy of the latent space to be the negative log of $\pi_{q_v}(q_v)$:
\begin{align*}
	U_h(q_h) = U_v(\psi(q_h)) = -\log \pi_{q_v}(\psi(q_h)) 
\end{align*}
Notice that this is not a re-parameterization since we will still use the potential energy function from the original space for our inference. If we use the density function of $q_h$ induced by $\phi(q_v)$ to be the potential energy, we will need to evaluate the volume change at each leapfrog step, which increases computational costs. As long as we could ensure detailed balance in the original space, which we will prove in later section, the MCMC proposal mechanism will be valid.

We also set the kinetic energy as follows:
\begin{align} \label{eq:K_h}
	K_h(p_h) = K_v(\psi(p_h)) = \psi(p_h)^TM^{-1}\psi(p_h)/2
\end{align}
Thus, we simulate the following Hamiltonian dynamics in the latent space:
\begin{align*}
	\dfrac{dq_{hi}}{dt} & = \dfrac{\partial K_h(p_h)}{\partial p_{hi}} \\
	\dfrac{dp_{hi}}{dt} &= -\dfrac{\partial U_h(q_h)}{\partial q_{hi}} 
\end{align*}

The evaluation of the gradient of the potential function with respect to $q_h$ can be calculated by the chain rule. For example, in the experiments discussed below, the decoder has one hidden layer with activation function $tanh$, and the connection to the output layer is linear. We can calculate the gradient function with respect to the latent variable $q_h$ as follows:
\begin{align} \label{eq:grad_U_h}
	\begin{split}
		\dfrac{\partial U_h(q_h)}{\partial q_{h}} & =  \left(\dfrac{\partial q_v}{\partial q_h}\right)^T \dfrac{\partial U_v(q_v)}{\partial q_{v}}
	\end{split}
\end{align}
where
\begin{align*}
	\begin{split}		
		\dfrac{\partial q_v}{\partial q_h} & = D_2 \mathrm{diag}(1-\tanh^2(D_1q_h +b_1)) D_1 \\
		\tanh(z) &= \dfrac{e^z-e^{-z}}{e^z+e^{-z}}, \quad \tanh'(z) =1-\tanh^2(z)
	\end{split}
\end{align*}
where $D_1$ and $D_2$ are the estimated weights of the decoder (Figure \ref{fig:auto-encoder}).
A detailed calculation of the gradient of $U(q_h)$ regarding a logistic regression example can be found in appendix. The resulting gradient evaluation is less expensive because of the much lower dimension.

The evaluation of $\frac{\partial K_h(p_h)}{\partial p_h}$ can be done in a similar way with $\frac{\partial K_h(p_h)}{\partial p_v} = M^{-1}p_v$ and $p_v = D_2 \tanh (D_1p_h + b_1)$, 
\begin{align} \label{eq:grad_K_h}
	\dfrac{\partial K_h(p_h)}{\partial p_h}  =& \{D_2 \mathrm{diag}(1-\tanh^2(D_1p_h +b_1)) D_1\}^T \cdot \nonumber  \\ 
	&M^{-1 } D_2 \tanh (D_1p_h + b_1) \\ 
	=& D_1^T \mathrm{diag}(1-\tanh^2(D_1p_h +b_1))  \cdot \nonumber \\ 
	& D_2^TM^{-1}D_2 \tanh (D_1p_h + b_1)  \nonumber
\end{align}
where $D_2^TM^{-1}D_2 $ can be pre-calculated.

In practice, the Hamiltonian dynamics is simulated using leapfrog steps.

\subsection{Proposal and correction}
\paragraph{Joint distribution of $(q_v, p_v)$} The density of $p_v$ is selected to be zero-mean Gaussian with a covariance $M$, corresponding to $K_v(p_v)$ defined in equation (\ref{eq:K_h}). We then have the canonical distribution of the original phase space: 
\begin{align} \label{eq:canonical}
	\pi_{q_v, p_v}(q_v, p_v) & \propto \exp( \log \pi_{q_v}(q_v)-  p_v^TM^{-1}p_v/2)
\end{align}

Notice the induced dynamics in the phase space $(q_v, p_v)$ is no longer Hamiltonian, and does not have the property of volume preservation as standard HMC. We hereby prove that the proposed HMC update will leave the canonical distribution for $q_v$ and $p_v$ (equation \ref{eq:canonical}) invariant, assuming that
\begin{enumerate}[i]
	\item the update is time reversible and thus symmetrical
	\item an appropriate volume correction term is added in the HMC acceptance probability
\end{enumerate}

\paragraph{Time reversibility} The proof is straightforward. Let $T_s$ represent the Hamiltonian dynamic in the latent space from the state $(q_h, p_h)$ at time $t$ to the state $(q_h^*, p_h^*)$ at time $t+s$. The reversibility of Hamiltonian dynamics indicates that:
\begin{align*}
	& T_s(q(t),p(t))  = (q(t+s), p(t+s)) \\
	& T_s(q(t+s), -p(t+s))  = (q(t), -p(t))
\end{align*} 
If we let $Q(q_h^*, p_h^* \vert q_h, p_h)$ represent the process of negating momentum, applying mapping $T_s$ and negating the momentum again,  and let $\Phi(q_v, p_v) = (\phi(q_v), \phi(p_v))$, $\Psi(q_h, p_h) = (\psi(q_h), \psi(p_h))$, our proposal can be denoted $Q' = \Psi \circ Q \circ \Phi$. We must have:
$$Q'(q_v^*, p_v^* \vert q_v, p_v) = Q'(q_v, p_v \vert q_v^*, p_v^*)$$ A detailed proof can be found in appendix.

\paragraph{Detailed Balance with Volume Correction} Following the proof in \cite{neal2011mcmc}, we can show that when accounting for volume change in the acceptance ratio, detailed balance holds for our proposed Metropolis update.

Consider partitioning the phase space $(q, p)$ into small regions $A_k$ with small volume $V$. Suppose by applying mapping $Q'$ to $A_k$, the image of $A_k$ becomes $B_k$. The $B_k$ will also partition the space due to reversibility, but has a different volume $V'$. We need to show detailed balance:
\begin{align*}
	P(A_i)T(B_j \vert A_i) = P(B_j)T(A_i \vert B_j) \quad \forall i, j
\end{align*}
Since when $i \neq j$, $T(B_j \vert A_i) = T(A_i \vert B_j) = 0$, we only consider when $ i = j \equiv k$:
\begin{align*}
	T(B_k \vert A_k) = Q'(B_k \vert A_k) \min(1, \dfrac{\exp(-H_{B_k})}{\exp(-H_{A_k})} \dfrac{V'}{V} )
\end{align*} 
See more details in appendix.

\begin{algorithm}[t]
	\caption{Auto-encoding HMC (AE-HMC)}
	\label{alg:auto-encoder_hmc}
	\begin{algorithmic}
		\State \textbf{Inputs}:
		\State \hspace{0.5cm}  encoder $\phi$, decoder $\psi$
		\State \hspace{0.5cm}  $U_v(q_v)$
		\State \hspace{0.5cm} $grad\_U_h(q_h)$ according to equation (\ref{eq:grad_U_h})
		\State \hspace{0.5cm} $grad\_K_h(p_h)$ according to equation (\ref{eq:grad_K_h})
		\State \hspace{0.5cm}  auto-encoder weights and biases $W, b$
		\State \hspace{0.5cm}  step size $\epsilon$, number of leapfrog steps $L$
		\State \hspace{0.5cm} current $q_v$
		\State
		\State Initialize $q_v^{(0)} =$ current $q_v$
		\State Sample momentum $p_v^{(0)} \sim \mbox{Normal}(0, M)$
		\State Set $q_h^{(0)} = \phi(q_v^{(0)})$
		\State Set $p_h^{(0)} = \phi(p_v^{(0)})$
		\For{ $i = 1 $ to $L$}
		\State $p_h^{(i-1/2)} = p_h^{(i-1)} - \epsilon/2 \cdot grad\_U_h(q_h^{(i-1)})$
		\State $q_h^{(i)} = q_h^{(i-1)} + \epsilon \cdot grad\_K_h(p_h^{(i-1/2)} )$
		\State $p_h^{(i)} = p_h^{(i-1/2)} - \epsilon/2 \cdot grad\_U_h(q_h^{(i)})$
		\EndFor 
		\State Calculate $q_v^{(L)} = \psi(q_h^{(L)})$
		\State Calculate $\rho = \exp(-H(q_v^{(L)}, p_v^{(L)}) + H(q_v^{(0)}, p_v^{(0)})) \cdot \begin{vmatrix}
		\dfrac{\partial(q_v^{(L)}, p_v^{(L)})}{\partial(q_v^{(0)}, p_v^{(0)})}
		\end{vmatrix}$ according to equation (\ref{eq:rho}) ($W, b$ will be needed accordingly)
		\State Sample $u \sim \mbox{Uniform}(0, 1)$
		\If {$u < \min(1, \rho)$}
		\State \Return $q_v^* = q_v^{(L)}$
		\Else
		\State \Return $q_v^* = $ current $q_v$
		\EndIf
	\end{algorithmic}
\end{algorithm}

\paragraph{Calculation of acceptance ratio}  For acceptance ratio $\alpha = \min(1, \rho)$, we have 
\begin{align*}
\begin{split}
	\rho = \exp(-H(q_v^*, p_v^*) +H(q_v, p_v) ) \begin{vmatrix}
		\dfrac{\partial(q_v^*, p_v^*)}{\partial(q_v, p_v)}
	\end{vmatrix} 
\end{split}
\\[2ex]
\begin{split}
	=\exp(-U_v(q_v^*)+U_v(q_v)-K_v( p_v^*) +K_v(p_v) )  \begin{vmatrix}
		\dfrac{\partial(q_v^*, p_v^*)}{\partial(q_v, p_v)}
	\end{vmatrix}
\end{split}	
\end{align*}
The determinant of $\dfrac{\partial(q_v^*, p_v^*)}{\partial(q_v, p_v)}$ is infeasible to evaluate. As shown in appendix, $\begin{vmatrix}
\dfrac{\partial(q_v^*, p_v^*)}{\partial(q_v, p_v)}
\end{vmatrix}$ can be approximated by $
\dfrac{\textrm{Vol} (q_v^*, p_v^*)}{\textrm{Vol}(q_h^*, p_h^*)}
\dfrac{\textrm{Vol} (q_h, p_h)}{\textrm{Vol}(q_v, p_v)}$. 

These two Jacobian matrices are not full rank, so we use the square root of its Gramian function $G(\cdot)$ to calculate the volume change (see appendix for more details). Thus, we have:
\begin{align} \label{eq:rho}
	\rho =& \exp(-U_v(q_v^*)+U_v(q_v)-K_v( p_v^*) + K_v(p_v)) \cdot \\
	& \sqrt{G(\dfrac{\partial (q_v^*, p_v^*)}{\partial (q_h^*, p_h^*)})}\sqrt{ G(\dfrac{\partial (q_h, p_h)}{\partial(q_v, p_v)})}
\end{align}

Algorithm \ref{alg:auto-encoder_hmc} summarizes the steps for a single iteration of AE-HMC.

\subsection{Approximate Bayes inference}
Given finite computational resources, in practice it would be more reasonable to keep a balance between accuracy of estimates and speed of computation. To do this, we can drop the volume adjustment. This way, instead of converging to the true distribution, our algorithm converges to an approximate but reasonably accurate distribution. In Bayesian analysis, if prediction is the ultimate goal, the resulting drop in the accuracy of estimating the posterior distribution might not lead to substantial deterioration in the prediction accuracy. Note that our method would still be preferable to those that only provide point estimates for predictions since it provides a reasonable estimate of prediction uncertainty. 

\paragraph{Illustration} 
To demonstrate that the approximate posterior distribution provided by AE-HMC can still capture high probability regions, we conduct a high-dimensional logistic regression experiment with simulated data.

In this experiment, we create a relatively challenging synthetic dataset for binary classification with 500 features from which 50 of them are highly correlated ($\rho=0.85$), and the rest are sampled from a standard normal distribution. In total, we simulate 550 data points for training and 150 for testing. We assume $\beta$'s have $N(0, 10^2)$ priors.

\begin{figure}[!t]
		\centering
		\includegraphics[height=7cm, width=8cm]{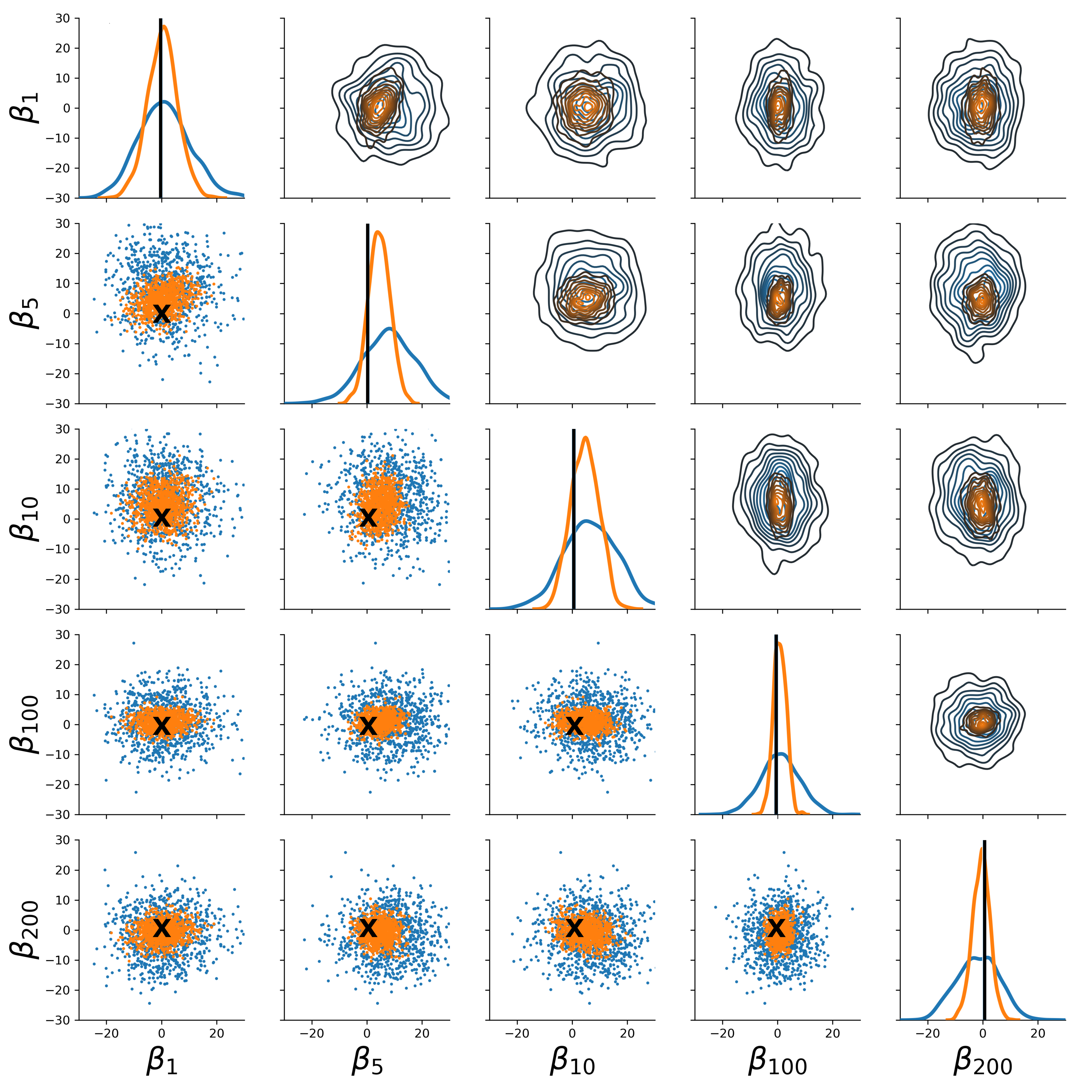}
		\caption{Comparing posterior distributions based on Standard HMC (blue) and Auto-encoding HMC (orange). As we can see, although the approximate distributions provided by AE-HMC tend to be more concentrated, they clearly cover the true parameter values (black).}
		\label{fig:lr_comparison}
\end{figure}

We first run the standard HMC to obtain 1000 samples, which are used to train the auto-encoder. For both standard HMC and AE-HMC, we tune the acceptance rate to be around 0.65 to 0.7, which is optimal in terms of computational efficiency \cite{beskos2013optimal}. 
We compare the results using $1000$ samples after the convergence has been reached. 


Figure \ref{fig:lr_comparison} shows the posterior distributions of five different $\beta$'s for both sampling methods. As we can see, our method still provides a reasonable approximation of the posterior distribution (at much lower computational cost as shown in the next section). Similar results are obtained for other parameters.  

\section{Experiments}
In this section, we evaluate the performance of our proposed method by comparing it to standard HMC. To this end, we will use two types of models: 1) high dimensional Bayesian logistic regression, and 2) high dimensional Bayesian inverse problem with Eliptic PDE.  

\subsection{High-dimensional Bayesian logistic regression}
Along with the synthetic dataset discussed above, we also examine our method based on three real datasets: CNAE-9, Optical Recognition of Handwritten Digits (both from the UCI Machine Learning Repository), and MNIST. For each dataset, we only focus on the first two groups for binary classification.

We use STAN \cite{pystan2018} and Keras \cite{chollet2015keras} with the TensorFlow backend to implement our method and compare its performance to the standard HMC algorithm implemented in STAN. 

We use an auto-encoder of three fully connected layers with linear activation where the dimension of the middle layer is roughly ten times smaller than the input layer. In the first step, we generate initial samples (from the warm-up stage) for training the auto-encoder using STAN. These samples consisted of approximately 10\% of the full HMC simulation. Then the auto-encoder is trained using these samples and its weights are extracted for custom implementation in STAN in order to generate samples from the latent space. Then, the samples are projected back to the original space followed by the accept-reject step. 

As we can see in Table \ref{tab:logistic_datasets}, our method could substantially improve the speed of Bayesian inference, while providing accuracy rates similar to those of standard HMC. 

\setlength{\tabcolsep}{3pt}
\begin{table}
\scriptsize
    \centering
    \begin{tabular}{llcccc}\toprule
    Method & Dataset & Synthetic & CNAE9 & Digits & MNIST\\\midrule
        & \#Parameters & 500 & 856 & 64 & 784\\ 
    HMC & Time & 11,077.8 & 3,793.3 & 865.7 & 33,300.8\\
        & Accuracy & 82\% & 98.3\% & 100\% & 99.6\%\\ \midrule
        & \#Parameters & 50 & 85 & 6 & 78\\
    AE-HMC & Time & 3,471.8 & 1,083.0 & 145.6 & 4,878.5\\
        & Accuracy & 82\%  & 98.3\% & 100\% & 99.2\% \\ \midrule
    Speed-up & & \textbf{3.2} & \textbf{3.5} & \textbf{5.9} & \textbf{6.8}\\    
    \bottomrule
    \end{tabular}

    \caption{Comparing accuracy rate (on test sets) and computational cost (time in second) of HMC and AE-HMC based on four different logisitic regression models. The number of parameters for AE-HMC shows the dimension of the latent space. Both methods use the same number of MCMC iterations.}
    \label{tab:logistic_datasets}
\end{table}

\subsection{High-dimensional Bayesian inverse problem with elliptic PDE}
Next, we examine the performance of our method using a more complex model --- Bayesian inverse problem. The model involves the following elliptic PDE defined on the unit square domain $\Omega=[0,1]^2$:
\begin{equation*}\label{eq:elliptic}
\begin{aligned}
-\nabla \cdot (k(s) \nabla p(s)) &= f(s), \; s\in \Omega \\
\langle k(s) \nabla p(s), \vec n(s) \rangle &= 0, \; s \in \partial\Omega \\
\int_{\partial\Omega} p(s) dl(s) &= 0
\end{aligned}
\end{equation*}
where $k(s)$ is the transmissivity field, $p(s)$ is the potential function, $f(s)$ is the forcing term,
and $\vec n(s)$ is the outward normal to the boundary.

\begin{figure}[b]
\centering
\includegraphics[scale=0.5]{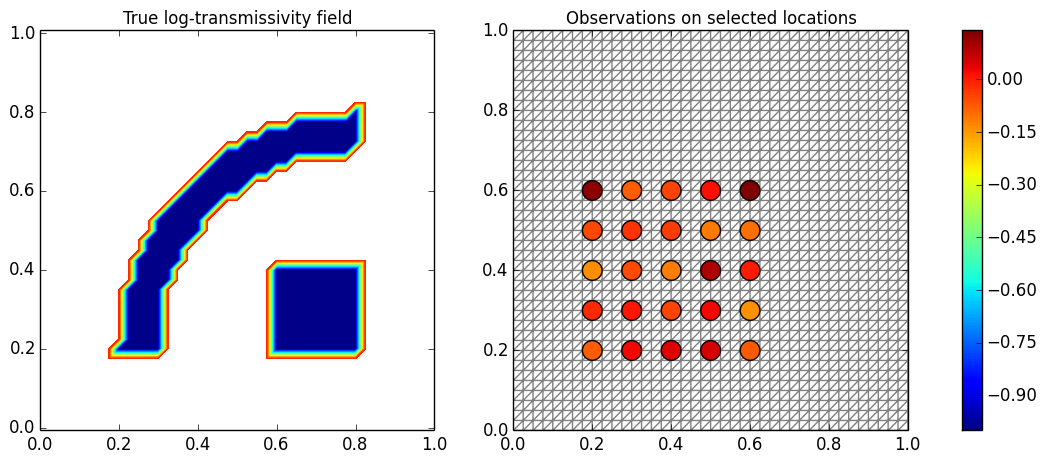}
\caption{True log-transmissivity field $u_0(s)$ (left), and $25$ observations on selected locations indicated by circles (right), with color indicating their values.}
\label{fig:truth_obs}
\end{figure}

To generate data, we construct a true transmissivity field $k_0(s)$ as shown on the left panel of the Figure \ref{fig:truth_obs}.
Partial observations are obtained by solving $p(s)$ on an $80\times 80$ mesh and then collecting at $25$ measurement sensors as shown by the circles on the right panel of the Figure \ref{fig:truth_obs}. 
The corresponding observation operator $\mathcal O$ yields the data
\begin{equation*}
y = \mathcal O p(s) + \eta, \quad \eta \sim \mathcal N(0, \sigma_\eta^2 I_{25})
\end{equation*}
where the signal-to-noise ratio is set at $\textrm{SNR}:=\max_s \{u(s)\}/\sigma_\eta=10$.

The inverse problem involves finding the transmissivity field $k(s)$ from the observations.
Bayesian approach endows a log-Gaussian prior for $k(s)$:
\begin{equation*}
k(s) = \exp(u(s)), \quad u(s) \sim \mathcal N(0, \mathcal C)
\end{equation*}
where the covariance operator $\mathcal C$ is defined through an exponential kernel function
\begin{align*}
\mathcal C: & \mathbb X \rightarrow \mathbb X, \; u(s) \mapsto \int c(s, s') u(s') ds', \\
 c(s, s') &= \sigma_u^2 \exp \left( -\frac{\Vert s- s' \Vert}{2s_0}\right), \,\textrm{for}\; s,s' \in \Omega
\end{align*}
with the prior standard deviation $\sigma_u=1.25$ and the correlation length $s_0=0.0625$ in the experiment.
Then, the problem reduces to sampling from the posterior of the log-transmissivity field $u(s)$, which becomes a vector of dimension over $6500$ after being discretized on $40\times 40$ mesh (with Lagrange degree $2$). See more details in \cite{cui16, LAN201971}

This is a very challenging sampling problem. To make the sampling rigorous in such a high dimensional space, we refer to the following pre-conditioned Crank-Nicolson (pCN) proposal,
which can be viewed as a variant of RWM \cite{Beskos2017, LAN201971}:
\begin{equation*}
q_v^*=\rho\,q_v+\sqrt{1-\rho^2}\,p_v\ , \quad p_v \sim \mathcal N(0, \mathcal C)
\end{equation*}
where $\rho = (1-\tfrac{h}{4})/(1+\tfrac{h}{4})$ with $h$ being the step size.
We follow the same procedure as AE-HMC to project $q_v$ to $q_h$ in the latent space of a much smaller dimension e.g. 600,
make a proposal $q_h^*$ based on the above proposal and finally map it back the original space. We refer to this modified version of our method as AE-pCN.

\setlength{\tabcolsep}{9pt}
\begin{table}[t]
\scriptsize
    \centering
    \begin{tabular}{llcc}\toprule
    Method & Mesh & $(20\times20)$ & $(40\times40)$  \\ \midrule
    
        & Parameters & 1681 & 6561\\
    pCN & Time & 4,368.7 & 15719.7\\
        & Log-likelihood & -7.47 & -7.46\\ \midrule
        
        & Parameters* &  441 & 441\\
    AE-pCN & Time & 1,008.6 & 1,038.3\\
    $(10\times10)$    & Log-Likelihood & -7.906 &  -7.915\\\midrule
    Speed-up & & \textbf{4.3}  & \textbf{15.1}\\ \bottomrule
    \end{tabular}

    \caption{Comparing accuracy (in terms of log-likelihood) and computational cost (time in second) of pCN and AE-pCN based on a high dimensional Bayesian inverse problem involving Eliptic PDE. The number of parameters for AE-HMC shows the dimension of the latent space. Both methods use the same number of MCMC iterations.}
    \label{tab:pde_results}
\end{table}

We run pCN on two mesh sizes $(20\times20)$ and $(40\times40)$ which  are reduced to a problem of size $(10\times10)$ using AE-pCN. We observe on Table \ref{tab:pde_results} a significant reduction on computation time but similar log-likelihood values using our proposed method. These results indicate an excellent trade-off between computational run time and the log-likelihood approximation.

\section{Discussion}
In this paper, we have proposed a new approach for approximating high dimensional probability distributions for fast, yet accurate Bayesian inference. Using synthetic and real data, we have shown that the resulting algorithm achieves a good balance between computational cost and posterior approximation.

There are some possible future directions worth pursuing. First, our method loses a nice property of standard HMC, namely, volume preservation. If a volume preserving embedding can be developed, it will allow for better approximation of posterior distribution. 

The computational saving mainly depends on the dimension of the latent space. However, reducing dimension empirically could lead to substantial loss of information. More work needs to be done to automatically determine the optimal size of the latent space. In addition, it is conceivable that some specific auto-encoder architectures and activation functions could provide better trade-offs between computational cost and accuracy.

Finally, note that our work can be extended to other MCMC algorithms using a similar framework (as shown in the previous section). Also, because our method focuses on high-dimensional problems with a large number of parameters, conceptually, it can be combined with some recent sampling algorithms that focus on problems with large sample sizes. 

\section*{Acknowledgement}
This work is supported by NSF grant DMS 1622490 and NIH grant R01 MH115697.

\appendix

\section*{Appendix}

\section{Calculating the gradient of $U_h(q_h)$ for Bayesian logistic regression} \label{appendix:gradient_lr}
Consider a logistic regression model  $y_i|{\bf X}_i, q_v \sim \mathrm{Bern}(\pi_i),\quad \pi_i=\frac{1}{1+\exp(-{\bf X}_i  q_v)}$, where ${\bf X}_i = (X_{i1}, X_{i2}, \cdots, X_{iD}), \quad q_v = (q_{v1},q_{v2}, \cdots, q_{vD})^T$. We set the prior to be zero mean Gaussian with unit variance.

The total Likelihood is $p(y \vert {\bf X}, q_v ) = \prod \limits_{i = 1}^N \left( \dfrac{1}{1+\exp (- {\bf X}_{i }q_v) } \right)^{y_i} \left( \dfrac{1}{1+\exp ({\bf X}_{i}q_v)} \right)^{1-y_i}$

Given that the decoder has one hidden layer and $tanh$ is used as the activation function, we have:  $$q_v = D_2 \tanh (D_1q_h + b_1).$$ Consider
\begin{align*}
	U_v(q_v) &= -\log(p(q_v)) - \log p(y \vert {\bf X}, q_v ) \\ 
	& = \dfrac{1}{2}q_v^Tq_v - \sum \limits_{i = 1}^N \left( y_i \log \dfrac{1}{1+\exp (- {\bf X}_{i}q_v)} + (1-y_i) \log \dfrac{1}{1+\exp ({\bf X}_{i}q_v)} \right) \\
	& =  \dfrac{1}{2}q_v^Tq_v - \sum \limits_{i = 1}^N y_i ({\bf X}_{i}q_v) + \sum \limits_{i = 1}^N \log (1+\exp ({\bf X}_{i}q_v)) \\
	\dfrac{\partial U_v(q_v)}{\partial q_v} & = q_v - {\bf X}^T_{D\times N} \left( y - \dfrac{1}{1+\exp (-{\bf X} q_v)}\right)_{N\times 1}\\
	& = q_v - \sum \limits_{i = 1}^N  {\bf X}_{i}^T y_i + \sum \limits_{i = 1}^N \dfrac{{\bf X}_{i}^T}{1+\exp (-{\bf X}_{i}q_v)}
\end{align*}
Thus, \begin{align}
	\begin{split}
		\dfrac{\partial U_h(q_h)}{\partial q_h} =& \dfrac{\partial U_v(q_v)}{\partial q_h}  \\
		= & \left(\dfrac{\partial q_v}{\partial q_h}\right)^T \dfrac{\partial U_v(q_v)}{\partial q_v} \\
		= & \left[ D_2 \mathrm{diag}(1-\tanh^2(D_1q_h +b_1)) D_1\right]^T \cdot \\
		   &  \left[ D_2 \tanh (D_1q_h + b_1) -  {\bf X}^T y + {\bf X}^T\dfrac{1}{1+\exp (-{\bf X}D_2\tanh(D_1q_h+b_1))}\right ] \\
		= &  D_1^T \mathrm{diag}(1-\tanh^2(D_1p_h +b_1)) \cdot \\
		   &  \left[ D_2^TD_2 \tanh (D_1q_h + b_1) - \sum \limits_{i = 1}^N ({\bf X}_{i}D_2)^T y_i+ \sum \limits_{i = 1}^N \dfrac{({\bf X}_{i}D_2)^T}{1+\exp (-{\bf X}_{i}D_2\tanh(D_1q_h+b_1))}\right ]
	\end{split}
\end{align}
where $D_2^TD_2$ and $X_iD_2$ can be pre-calculated. Notice that dim$(X_iD_2) <<$ dim$(X_i)$.

\section{Time reversibility} \label{appendix:time_reverse}
Given that $\psi$ is the inverse map of $\phi$, and define function $h(q, p) = (q, -p)$ we have
\begin{align*}
	\Psi(Q(\Phi(q_{v}, p_{v}))) &= \Psi(Q(\phi(q_{v}), \phi(p_{v}))) \\
	&= \Psi(Q(q_{h}, p_{h})) \\
	&= \Psi(h\circ T_s \circ h(q_{h}, p_{h})) \\
	&= \Psi(h\circ T_s(q_{h}, -p_{h})) \\
	&= \Psi(h(q_{h}^*, -p_{h}^*)) \\
	&= \Psi(q_{h}^*, p_{h}^*) \\
	&=  (\psi(q_{h}^*), \psi(p_{h}^*)) \\
	&=  (q_{v}^*, p_{v}^*) 
\end{align*}
and 
\begin{align*}
	\Psi(Q(\Phi(q_{v}^*, p_{v}^*))) &= \Psi(Q(\Psi^{-1}(q_{v}^*, p_{v}^*))) \\
	&= \Psi(Q(\psi^{-1}(q_{v}^*), \psi^{-1}(p_{v}^*))) \\
	&= \Psi(Q(q_{h}^*,p_{h}^*)) \\
	&= \Psi(h\circ T_s \circ h(q_{h}^*,p_{h}^*)) \\
	&= \Psi(h\circ T_s(q_{h}^*,-p_{h}^*)) \\
	&= \Psi(h(q_{h},-p_{h})) \\
	&= \Psi(q_{h}, p_{h}) \\
	&= \Phi^{-1}(q_{h}, p_{h}) \\
	&= (\phi^{-1}(q_{h}), \phi^{-1}(p_{h})) \\
	&=  (q_{v}, p_{v}) 
\end{align*}
Thus  $Q' = \Psi \circ Q \circ \Phi$ is symmetric.

\section{Detailed Balance with Volume Correction}
Following the proof in \cite{neal2011mcmc}, we can show that when accounting for volume change in the acceptance ratio, detailed balance holds for our proposed Metropolis update.

Consider partitioning the phase space $(q, p)$ into small regions $A_k$ with small volume $V$. Suppose by applying mapping $Q'$ to $A_k$, the image of $A_k$ becomes $B_k$. The $B_k$ will also partition the space due to reversibility, but has a different volume $V'$. We need to show detailed balance:
\begin{align*}
	P(A_i)T(B_j \vert A_i) = P(B_j)T(A_i \vert B_j) \quad \forall i, j
\end{align*}
Since when $i \neq j$, $T(B_j \vert A_i) = T(A_i \vert B_j) = 0$, we only consider when $ i = j \equiv k$. Let \begin{align*}
	T(B_k \vert A_k) = Q'(B_k \vert A_k) \min(1, \dfrac{\exp(-H_{B_k})}{\exp(-H_{A_k})} \dfrac{V'}{V} )
\end{align*} 
Then, we can write $P(A_k)T(B_k \vert A_k)$ as
\begin{align*}
	 &  V \exp(-H_{A_k}) Q'(B_k \vert A_k) \min(1, \dfrac{\exp(-H_{B_k})}{\exp(-H_{A_k})} \dfrac{V'}{V} ) \\
	=& Q'(B_k \vert A_k) \min(V \exp(-H_{A_k}), V' \exp(-H_{B_k}) ) \\
	=& Q'(A_k \vert B_k) \min(V \exp(-H_{A_k}), V' \exp(-H_{B_k}) ) \\
	=& V' \exp(-H_{B_k}) Q'(B_k \vert A_k) \min(\dfrac{\exp(-H_{A_k})}{\exp(-H_{B_k})} \dfrac{V}{V'}, 1) \\
	=& P(B_k)T(A_k \vert B_k) 
\end{align*}
The volume correction term $\dfrac{V'}{V}$ is simply the determinant of the Jacobian matrix $\begin{vmatrix}
\dfrac{\partial(q_v^*, p_v^*)}{\partial(q_v, p_v)}
\end{vmatrix} $.

\section{Approximating volume correction term}
\label{appendix:approximate_determinant}
Following \cite{neal2011mcmc}, let's consider a one dimensional example. For mapping
\begin{align*}
	T_{\delta}(q,p) = \begin{bmatrix}
		q \\
		p
	\end{bmatrix} + \delta \begin{bmatrix}
		dq/dt \\
		dp/dt
	\end{bmatrix} + O(\delta^2)
\end{align*} 
The Jacobian matrix:
\begin{align*}
	B_{\delta} = \begin{bmatrix}
		1+\delta \dfrac{\partial^2H}{\partial q \partial p} & \delta \dfrac{\partial^2H}{\partial p^2} \\
		-\delta \dfrac{\partial^2H}{\partial q^2} & 1-\delta \dfrac{\partial^2H}{\partial p \partial q}
	\end{bmatrix} + O(\delta^2)
\end{align*}
Consider a $3 \times 2$ matrix A and $2 \times 3$ matrix C:
\begin{align*}
	\begin{bmatrix}
		a_{11} & a_{12} \\
		a_{21} & a_{22} \\
		a_{31} & a_{32}
	\end{bmatrix} \left \{ \begin{bmatrix}
		1+\delta \dfrac{\partial^2H}{\partial q \partial p} & \delta \dfrac{\partial^2H}{\partial p^2} \\
		-\delta \dfrac{\partial^2H}{\partial q^2} & 1-\delta \dfrac{\partial^2H}{\partial p \partial q}
	\end{bmatrix} + O(\delta^2) \right\} \begin{bmatrix}
		c_{11} & c_{12} & c_{13} \\
		c_{21} & c_{22} & c_{23}
	\end{bmatrix} 
\end{align*}
It gives a $3 \times 3$ matrix with element $(i,j)$ to be $$a_{i1}c_{1j} + a_{i2}c_{2j} + \delta(a_{i1}c_{1j} \dfrac{\partial^2H}{\partial q \partial p} - a_{i2}c_{1j} \dfrac{\partial^2H}{\partial p^2} + a_{i1}c_{2j}\dfrac{\partial^2H}{\partial p^2} - a_{i2}c_{2j}\dfrac{\partial^2H}{\partial p \partial q}) + O(\delta^2)$$

We could show that $$\det(AB_{\delta}C) = \det(AC) + O(\delta^2)$$

The result can be generalized to higher dimensions.

Now let's denote $z = (q, p)$, $A_i = \dfrac{\partial z_v^i}{\partial z_h^i} $, $B_{\delta} =  \dfrac{\partial z_h^i}{\partial z_h^{i-1}}$, $C_i = \dfrac{\partial z_h^i}{\partial z_v^i} $. We have:
\begin{align*}
	\begin{split}
		\det(\dfrac{\partial z_v^L}{\partial z_v^0}) &= \det(\dfrac{\partial z_v^L}{\partial z_h^L} \dfrac{\partial z_h^L}{\partial z_h^{L-1}} \dfrac{\partial z_h^{L-1}}{\partial z_v^{L-1}}  \cdots \dfrac{\partial z_v^1}{\partial z_h^1} \dfrac{\partial z_h^1}{\partial z_h^0} \dfrac{\partial z_h^0}{\partial z_v^0}) \\
		&= \det(A_L B_{\delta} C_{L-1} A_{L-1} \cdots A_1 B_{\delta} C_0) \\
		&= \det(A_L B_{\delta} C_{L-1} ) \det(A_{L-1} B_{\delta} C_{L-2} ) \cdots \det(A_1 B_{\delta} C_0 ) \\
		&= (\det(A_LC_{L-1} ) + O(\delta^2)) (\det(A_{L-1} C_{L-2} ) + O(\delta^2)) \cdots (\det(A_1 C_0 ) + O(\delta^2)) \\
		&=  \dfrac{Vol( z_v^L)}{Vol(z_h^L)} \dfrac{Vol(z_h^0)}{Vol( z_v^0)} + O(\delta) \\
		& \rightarrow \dfrac{Vol( z_v^L)}{Vol(z_h^L)} \dfrac{Vol(z_h^0)}{Vol( z_v^0)} \mbox{ as } \delta \rightarrow 0
	\end{split}
\end{align*}

For matrices $A_L$ and $C_0$, the number of vectors are less than the dimension of the ambient space. We could use the square root of the gramian function of the matrix to calculate $k$-volume in $n$-space where $k < n$. In particular, for $k$ linearly independent vectors $v_1, \cdots,v_k$, the gramian function is $G(v1,...,vk) = det(M^TM)$ where $M = (v_1, \cdots,v_k) $. The volume of the parallelepiped with the vectors  is calculated by:
\begin{align*}
	Vol(v1,...,vk) = \sqrt{det(M^TM)}
\end{align*}

\bibliographystyle{unsrt}

\end{document}